\documentclass[11pt,a4paper]{article}

\usepackage[T1]{fontenc}
\usepackage[utf8]{inputenc}
\usepackage{newtxtext,newtxmath}
\usepackage[margin=25mm]{geometry}
\usepackage{microtype}
\usepackage{graphicx}
\usepackage{booktabs}
\usepackage{tabularx}
\usepackage{array}
\usepackage{siunitx}
\usepackage{enumitem}
\usepackage{caption}
\usepackage{float}
\usepackage{xcolor}
\usepackage{tikz}
\usetikzlibrary{arrows.meta,positioning,shapes.geometric,fit,calc,backgrounds}
\usepackage[hidelinks]{hyperref}
\usepackage[nameinlink,noabbrev]{cleveref}

\definecolor{AcademicBlue}{HTML}{3568A8}
\definecolor{AcademicInk}{HTML}{243447}
\definecolor{ReviewRed}{HTML}{A33A3A}
\usepackage[numbers,sort&compress]{natbib}
\setlist{leftmargin=*,itemsep=2pt,topsep=4pt}

\title{\textbf{A Hybrid Security Framework for Mini‑Programs: Visual UI Compliance and Network Risk Assessment}}

\author{
  Panpan Shen\\
  Hainan University, Haikou, China\\
  2949953001@qq.com
  \and
  Lei Xie\\
  Hainan University, Haikou, China\\
  xielei@hainanu.edu.cn
  \and
  Xiaoqi Li\\
  Hainan University, Haikou, China\\
  csxqli@ieee.org
}

\date{}

\begin{document}

\maketitle

\begin{abstract}
With the continuous development of the WeChat ecosystem, WeChat Mini Programs, due to their advantages of not requiring installation, using little memory, and being ready to use instantly, have seen a surge in user numbers and have now become an indispensable service carrier in mobile internet. However, as Mini Programs rapidly became popular, issues regarding the compliance of their interface interaction design and the safety of operational behavior have become increasingly apparent. Many Mini Programs have problems such as clickable buttons and icons not being standard in size, or ad pop-ups and payment entrances being placed in a way that is easy to misclick. The close or cancel buttons are often too small or hidden, making it easy to accidentally click on ads or payment content, and difficult to accurately click the cancel button. This can result in involuntary payments or being redirected to illegal pages, causing unnecessary financial losses and seriously harming users' property security and legal rights.  To address the above issues, this article develops a detection program to check the position and size of various icons and buttons in Mini Programs, and analyze whether redirected links fall within a safe range. YOLOv8 is used to identify various buttons in images, displaying the corresponding icon and its data based on the mouse click position. Violations are flagged and recorded. At the same time, mitmproxy is used to capture relevant data requests generated during clicks, analyzing the safety of redirections, and presenting key information for user observation.
\end{abstract}


\section{Introduction}

WeChat Mini-Programs\cite{wang2025miniscope} have rapidly evolved into a pivotal service paradigm within the mobile internet ecosystem\cite{zhang2021measurement}. By eliminating installation overhead, minimising memory consumption, and enabling instant launch, they have attracted an enormous and still-growing user base\cite{yang2025understanding}. Statistics indicate that by 2022, WeChat Mini-Programs had already exceeded 450 million daily active users, with over three million registered developers and an annual transaction scale reaching trillions of yuan. Industry statistics show that by the end of 2025, the number of cross-border and overseas interactions triggered by WeChat Mini-Programs exceeded 5 billion. Mini-Program \cite{mohammadi2025security} services have been deployed across 100 countries and regions worldwide, while WeChat cross-border payment capabilities cover 78 countries and regions with support for 36 types of currencies. Transaction volume completed via Mini-Programs\cite{yang2026real} in the second half of 2025 also saw a year-on-year growth of over 70\%. These figures collectively demonstrate that Mini-Programs have been deeply integrated into daily life scenarios ranging from food delivery and online shopping to mobility services, medical care and public utility payments\cite{wang2024smart}.

Despite their convenience and strong platform endorsement--including mandatory pre-release security reviews\cite{yang2026rise} and default HTTPS communication--the rapid proliferation of Mini-Programs has brought forth new safety and usability concerns. A substantial number of Mini-Programs exhibit non-standardised interactive controls: clickable buttons and icons are frequently rendered at undersized dimensions; advertisement pop-ups and payment entry points are placed in regions susceptible to unintentional touches; and close or cancel buttons are often deliberately miniaturised and tucked into inconspicuous positions\cite{2025Penetrating}. These design irregularities directly lead to inadvertent ad clicks, unintentional payment confirmations, and unwanted redirections to questionable web pages, thereby causing economic losses and undermining users' property security and legitimate rights\cite{2025Exploring}.

The research community has devoted considerable effort to Mini-Program security\cite{wang2022characterizing}. Early investigations focused on architectural vulnerabilities. Deng identified two prevalent categories of counterfeit Mini-Programs--those using visually similar icons and those mimicking official names via homophones or visually similar characters--and proposed a detection method combining code signatures and network behaviour\cite{deng2023detect}. Baskaran et al. examined code-level weaknesses and revealed that developer credentials, especially secret keys, are often hard-coded within the program package, creating a significant attack surface for credential leakage\cite{baskaran2023measuring,zhang2023don}. Subsequent work expanded the scope to privacy and data-flow risks. Zheng conducted both static and dynamic analyses on a large corpus of Mini-Programs, finding that over 60\% suffered from hard-coded API keys, missing HTTPS enforcement, inadequate signature verification, or local caching of sensitive user data such as phone numbers\cite{han2023systematic,zhao2023potential,zhang2023small}. Huang et al. further highlighted the abusive use of the web-view component to load external pages, effectively bypassing the platform's content review mechanism and opening doors to supply-chain poisoning and phishing attacks\cite{zhang2022identity,yang2022cross}.

In response to these threats, several automated analysis toolkits have been developed\cite{liu2020industry}. Meng implemented a pipeline that performs sensitive-API call tracing, hard-coded key extraction, and HTTPS configuration validation on decompiled code. Wemint employs static taint analysis to track sensitive data flows\cite{meng2023wemint,wang2023taintmini,li2023minitracker}, while more recent studies have leveraged large language models to detect real-world data leakage incidents\cite{chen2025whiskey}. Some researchers have combined static and dynamic methods to efficiently verify the integrity of data transmission paths\cite{wang2024miniscope,wang2024minichecker}. Meanwhile, platform regulators have progressively updated their guidelines--most notably, the \textit{WeChat Developer Code of Conduct} and \textit{Privacy Protection Specification}--mandating HTTPS for all network requests, requiring user-initiated authorisation for sensitive APIs, and enforcing the principle of minimum necessary data collection\cite{wang2023usage,wang2024you,li2024identifying}.

Nevertheless, a conspicuous gap remains in the current landscape of Mini-Program security testing: the lack of a dedicated visual-oriented and user-interactive inspection mechanism\cite{li2025interaction}. Existing UI-centric testing methodologies predominantly rely on random-event generation (e.g., Monkey-style stress tests) to evaluate crash resilience rather than to assess element-level compliance with human-factor design standards\cite{doshi2025phishhunter}. On the other hand, code-based security audits, while thorough at the logical level, ignore the spatial and behavioural attributes of interface elements that directly affect user actions. Furthermore, no prior work has integrated real-time screen-based icon/button inspection with network-request risk assessment in a single, interactive diagnostic tool\cite{zhang2025demystifying}.

To bridge this gap, this paper presents a hybrid security detection framework for WeChat Mini-Programs. The proposed system combines visual element recognition using YOLOv8 with network traffic analysis via mitmproxy, enabling both static UI compliance checking and dynamic behavioural risk prediction. The main contributions of this work are summarised as follows:
\begin{enumerate}[label=(\arabic*)]
    \item We design and implement a real-time screen-capture and preview module that eliminates self-window occlusion via Win32gui, ensuring that the detection model receives a clean, full-screen input image.
    \item We adopt YOLOv8--a state-of-the-art one-stage object detector--to identify and localise interactive UI elements (buttons, icons, close controls) with high precision. The detection is event-driven by mouse clicks using the Pynput library, avoiding continuous resource consumption.
    \item We establish a compliance judgement rule based on the human-centric $44\times44$ CSS-pixel threshold derived from the WCAG 2.1 standard and the WeUI Design Guidelines. Detected non-compliant elements are automatically highlighted with a red bounding box, a green $44\times44$ reference square is overlaid, and a timestamped screenshot is archived for evidence.
    \item We integrate mitmproxy to capture outbound data packets triggered by user clicks, and feed extracted features (URL length, IP presence, subdomain count, suspicious keywords, HTTPS usage, special-character frequency, and URL entropy) into a machine-learning risk classifier, thereby supplementing the visual inspection with network-level threat assessment.
\end{enumerate}

\section{Background}
\label{sec:background}

\subsection{Mini-Program Architecture and Security Implications}

WeChat Mini-Programs adopt a dual-threaded architecture that separates the rendering layer and the logic layer\cite{lu2020demystifying}. The rendering layer, implemented using WXML (WeiXin Markup Language) and WXSS (WeiXin Style Sheets), governs the visual layout, size, colour, and overall presentation of interface elements. The logic layer, written in JavaScript, handles business logic, page navigation, data exchange, and event responses\cite{wang2025wechat}. This separation enables efficient and asynchronous interaction but also introduces distinct attack surfaces: the rendering layer may be manipulated to obscure or misrepresent UI components, while the logic layer may contain insecure API calls or improperly validated data flows\cite{zhang2023minible}. A clear understanding of this architecture is essential for designing a detection system that targets both visual irregularities and suspicious behavioural patterns\cite{duru2025smart}.

\begin{figure}[H]
  \centering
  \resizebox{0.8\textwidth}{!}{\includegraphics{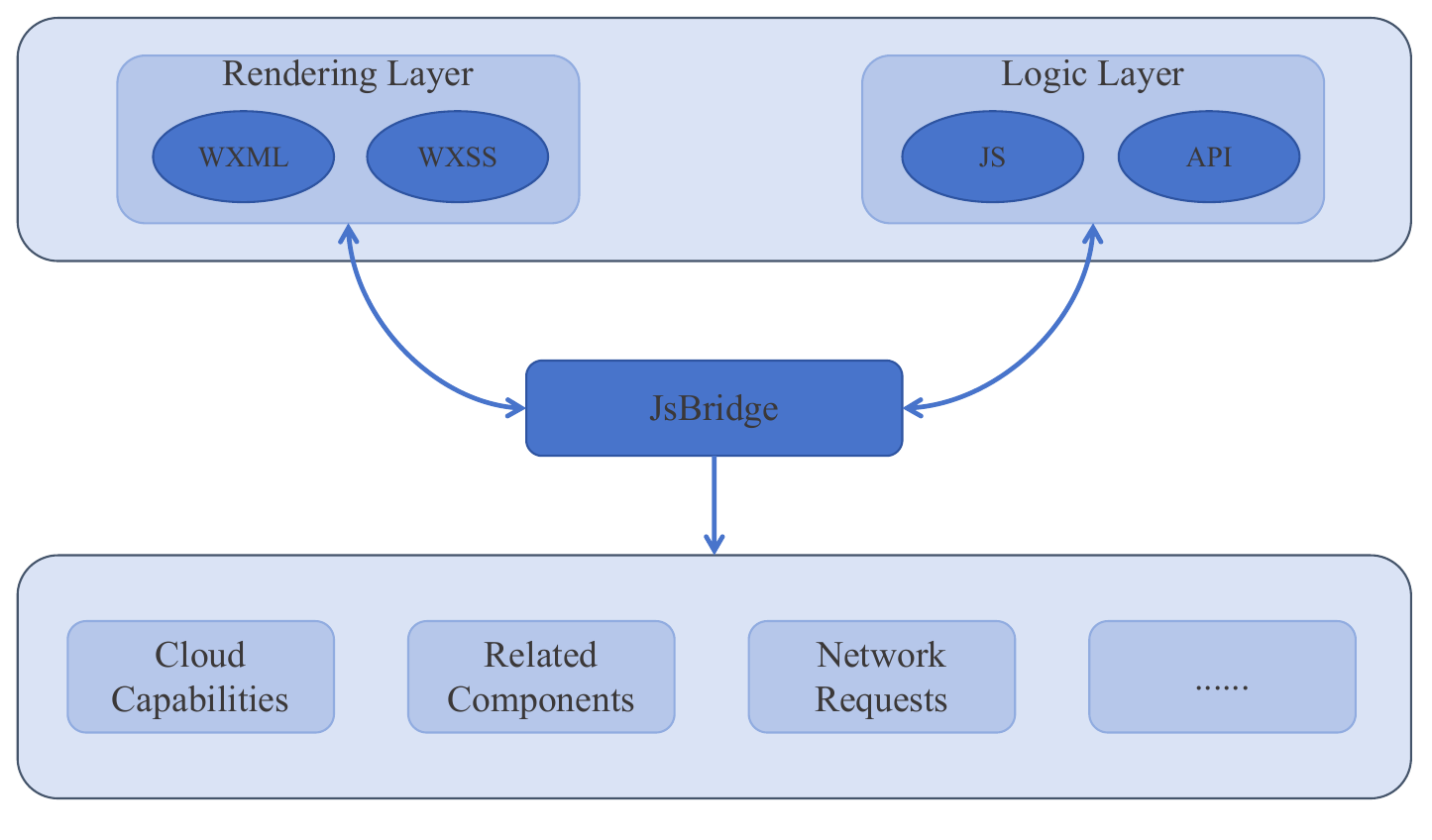}}
  \caption{Layered architecture of a WeChat Mini-Program and its JSBridge-mediated services.}
  \label{fig:architecture}
\end{figure}

\subsection{YOLOv8 for UI-Element Detection}

\begin{figure}[H]
 \centering
  \resizebox{\textwidth}{!}{\includegraphics{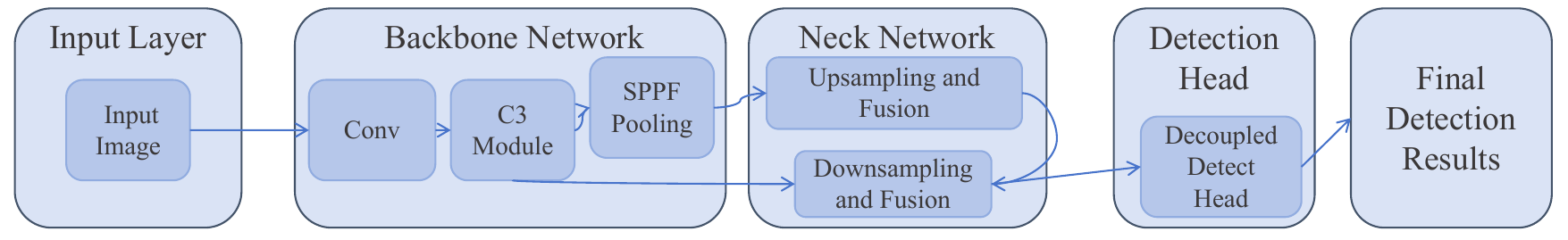}}
  \caption{Simplified YOLOv8 processing pipeline used for UI-element detection.}
  \label{fig:yolov8}
\end{figure}

The core visual recognition engine of the proposed detection system is YOLOv8 (You Only Look Once, version 8), a state-of-the-art one-stage object detection algorithm open-sourced by Ultralytics\cite{redmon2020you}. Inheriting and upgrading the advantages of previous YOLO series algorithms\cite{shayan2024gui}, YOLOv8 achieves higher detection accuracy and faster inference speed. It also provides multiple model sizes including nano, small, medium, large and extra-large to adapt to diverse computing resource constraints\cite{zhang2024dark}. The processing stages used in the manuscript are shown in \cref{fig:yolov8}.
Different from early visual testing tools like Sikuli that depend on rigid template matching and fail when icons change colour or aspect ratio, YOLOv8 extracts hierarchical feature representations via deep convolutional neural networks\cite{li2024exploring}. It exhibits strong robustness against changes in object scale, lighting conditions and partial occlusion. This characteristic is essential for Mini-Program detection, as UI components developed by different merchants differ greatly in appearance but share identical functional logic\cite{selcuk2023comparison}. For instance, the close button can be an "×" mark, left arrow or text label, all of which carry the same closing function\cite{2025Blockchain}.

YOLOv8 converts object detection into a regression task\cite{shi2025skeleton}. For each grid cell on the output feature map, the model predicts bounding box coordinates (centre $x$, centre $y$, width, height) and category probabilities. A composite indicator is used to screen positive samples, which is defined as:
\[
t = s^{\alpha} \times u^{\beta}
\]
where $s$ stands for classification confidence, $u$ denotes the Intersection over Union (IoU) between the predicted box and ground-truth box, and hyperparameters $\alpha$ and $\beta$ are generally set to 0.5 and 0.6 respectively. The candidate box with the maximum $t$ value is reserved as the final detection result to suppress false positive samples effectively.

In terms of bounding box regression, YOLOv8 adopts a coordinate-based regression strategy rather than the classic anchor offset method. Given the feature map coordinate $(x,y)$ and four distances from this point to the four edges of the target (left $l$, right $r$, top $t$, bottom $b$), the complete bounding box coordinate can be deduced as:
\[
(x-l,\; x+r,\; y-t,\; y+b)
\]
This structure can automatically adapt to screens with different resolutions and aspect ratios, without manually adjusting anchor parameters for various devices.

A compound loss function is used to train the YOLOv8 model and optimise positioning and classification effects simultaneously:
\[
L = \lambda_1 \cdot L_{\text{CIoU}} + \lambda_2 \cdot L_{\text{DFL}} + \lambda_3 \cdot L_{\text{BCE}}
\]
where $L_{\text{CIoU}}$ represents Complete IoU loss for bounding box regression, which comprehensively considers distance, overlapping area and aspect ratio matching degree\cite{zheng2020distance}; $L_{\text{DFL}}$ is Distribution Focal Loss for optimising the boundary probability distribution of detection boxes; $L_{\text{BCE}}$ denotes binary cross-entropy loss for classification tasks. As an alternative, Varifocal Loss $L_{\text{VFL}}$ can replace $L_{\text{BCE}}$ and assign higher weights to hard samples to improve detection performance on difficult targets.

\subsection{Screen Capture and Display}

PyAutoGUI provides cross-platform mouse and keyboard automation together with screen capture through \texttt{pyautogui.screenshot()}. Each captured frame is converted to an RGB NumPy array of shape $(h,w,3)$ before being passed to YOLOv8\cite{lu2025cloud}.
The general cross-platform control path\cite{wang2023one} represented in the source figure is redrawn in figure 3.

\begin{figure}[H]
  \centering
  \resizebox{0.5\textwidth}{!}{\includegraphics{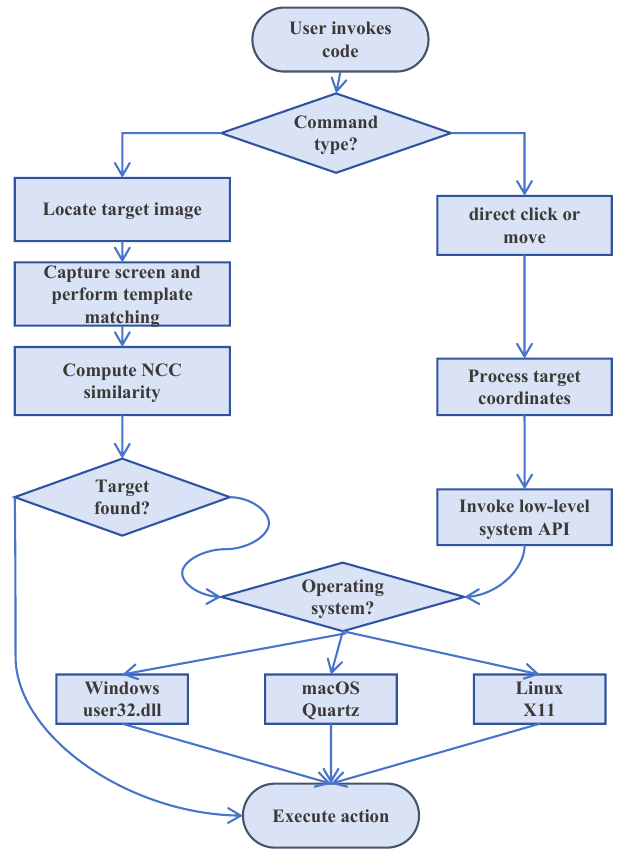}}
  \caption{Cross-platform image-location and direct-coordinate action flow.}
  \label{fig:cross-platform}
\end{figure}

A screenshot may include the detector's own window, producing picture-in-picture interference. On Windows, the system therefore uses Win32gui \cite{sun2025kubeguard}to enumerate top-level windows with \texttt{EnumWindows()}, identify its own window by title, obtain its rectangle with \texttt{GetWindowRect()}, and mask that region. This step provides a clean view of the Mini-Program under test.

\subsection{Event-Driven Interaction with Pynput}

To avoid continuous screen polling that wastes CPU resources and leads to interface stuttering, the entire detection process is only activated upon mouse click events. This work adopts Pynput, a low-level input monitoring library, to capture mouse clicks, movements and scrolling behaviours. Compared with the high-level simulation interfaces provided by PyAutoGUI, Pynput delivers better real-time response and more precise event control capability\cite{gao2026mind}.

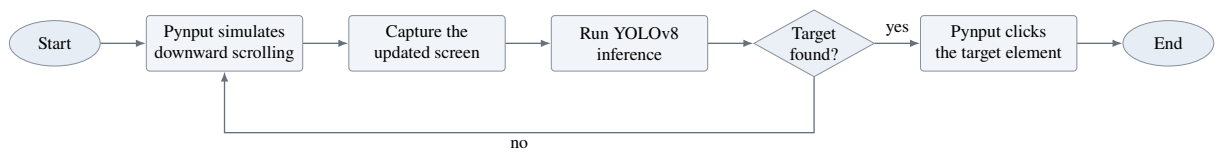
\begin{figure}[H]
  \centering
  \resizebox{\textwidth}{!}{\begin{tikzpicture}[
  font=\small,
  node distance=8mm and 9mm,
  flow/.style={-{Latex[length=2mm]}, draw=AcademicInk!70, line width=0.75pt},
  process/.style={draw=AcademicInk!55, rounded corners=2pt, fill=AcademicBlue!7,
    align=center, minimum width=31mm, minimum height=11mm},
  decision/.style={draw=AcademicInk!55, diamond, aspect=1.8, fill=AcademicBlue!10,
    align=center, inner sep=1.2pt},
  terminal/.style={draw=AcademicInk!55, ellipse, fill=AcademicBlue!12,
    align=center, minimum width=18mm, minimum height=9mm}
]
  \node[terminal] (start) {Start};
  \node[process, right=of start] (scroll) {Pynput simulates\\downward scrolling};
  \node[process, right=of scroll] (capture) {Capture the\\updated screen};
  \node[process, right=of capture] (infer) {Run YOLOv8\\inference};
  \node[decision, right=of infer] (found) {Target\\found?};
  \node[process, right=of found] (click) {Pynput clicks\\the target element};
  \node[terminal, right=of click] (end) {End};

  \draw[flow] (start) -- (scroll);
  \draw[flow] (scroll) -- (capture);
  \draw[flow] (capture) -- (infer);
  \draw[flow] (infer) -- (found);
  \draw[flow] (found) -- node[above] {yes} (click);
  \draw[flow] (click) -- (end);
  \draw[flow] (found.south) -- ++(0,-11mm) -| node[pos=0.25, below] {no} (scroll.south);
\end{tikzpicture}}
  \caption{Event-driven target search and click workflow implemented with Pynput.}
  \label{fig:pynput}
\end{figure}

Once a mouse click is captured, the system extracts the corresponding coordinate values and runs the YOLOv8 inference model based on the latest cleaned screen snapshot. After detection, the system screens all identified UI components and selects the one nearest to the click coordinate, then displays its category, pixel size and compliance state correspondingly. Moreover, a configurable cooldown timer is configured for scroll events captured by Pynput. This setting prevents repeated triggering from a single scroll operation, cuts down redundant calculations and stabilises the final statistical results.
\cref{fig:pynput} shows the corresponding event flow.

\subsection{Image-Evidence Retention}

A distinctive feature of the proposed framework lies in the automatic storage of violation evidence. When the detected component (mostly close, cancel or confirmation buttons) fails to meet the specification, namely its size is smaller than the $44\times44$ CSS-pixel threshold, the system executes a four-stage evidence saving pipeline as follows:
\begin{enumerate}[label=(\arabic*)]
    \item \textit{Trigger Validation}: Verify that the clicked UI element is a non-compliant control.
    \item \textit{Full-screen Capture}: Obtain the complete screen image by calling \texttt{pyautogui.screenshot()}.
    \item \textit{Image Annotation}: Render a red bounding box to enclose the irregular element and a red dot to mark the click position. A standard $44\times44$ green reference frame is added below the target region, along with text labels recording the violation category and actual pixel dimensions.
    \item \textit{Local Storage}: Export the annotated snapshot to a designated local folder in PNG format with a timestamp filename, which balances image clarity and disk space consumption.
\end{enumerate}

The proposed evidence archiving module supports subsequent review and program debugging, and delivers intuitive visual hints for both developers and ordinary users.

\subsection{Network-Traffic Analysis with mitmproxy}

Image-based inspection cannot determine whether a visually compliant control triggers a malicious or high-risk request. The system therefore integrates mitmproxy, a man-in-the-middle proxy, to intercept and analyse outbound HTTP and HTTPS traffic generated by Mini-Program operations. A whitelist of trusted domains, including \texttt{weixin.qq.com}, \texttt{qq.com}, \texttt{wechat.com}, \texttt{qcloud.com}, and \texttt{wx.tenpay.com}, filters routine platform traffic. The remaining requests are represented by the features in \cref{tab:url-features}.The classifier returns a risk score that complements the YOLOv8-based compliance result, covering both the on-screen presentation and the network-level behaviour of the clicked element.
\begin{table}[t]
  \centering
  \caption{Features used for URL-risk assessment.}
  \label{tab:url-features}
  \begin{tabular}{@{}p{0.28\textwidth}p{0.68\textwidth}@{}}
    \toprule
    \textbf{Feature} & \textbf{Description} \\
    \midrule
    URL length & Total number of characters in the request URL. \\
    IP-address presence & Whether the host is an IP literal, which may indicate evasion. \\
    Subdomain count & Number of dot-separated labels in the hostname. \\
    Suspicious-keyword count & Occurrences of terms such as \texttt{login}, \texttt{pay}, \texttt{transfer}, \texttt{wallet}, and \texttt{verify}. \\
    HTTPS use & Boolean indicator of whether TLS is used. \\
    Special-character count & Frequency of non-alphanumeric symbols such as \%, \&, $=$, and $+$. \\
    URL entropy & Shannon entropy of the URL string; a high value may indicate obfuscation. \\
    \bottomrule
  \end{tabular}
\end{table}

\section{System Design and Methodology}
\label{sec:method}

\subsection{Module and Layer Architecture}

The proposed detection system adopts a three-tier architectural design, which is divided into the data layer, business logic layer, and interaction layer in accordance with functional responsibilities and data transmission sequence. This layered structure achieves clear separation of concerns and effectively enhances the maintainability and scalability of the system.

\begin{enumerate}[label=(\arabic*)]
    \item \textit{Data layer}: Serving as the persistent data foundation of the entire system, this layer consists of two sub-modules for structured data management. The image storage module saves annotated violation screenshots into an exclusive local directory. To ensure file uniqueness and convenient retrieval, each screenshot is named with a timestamp and stored in PNG format, which balances image clarity and storage overhead. The module also supports automatic path creation to eliminate runtime errors caused by missing folders. The operation log module systematically records user interactions including mouse clicks and scroll events, providing traceable and structured data for subsequent statistical analysis, performance evaluation and violation investigation.
    \item \textit{Business logic layer}: As the core execution unit of the system, this layer integrates seven functional modules that collaborate through event-driven mechanisms and data transmission to complete the full detection workflow. Specifically, the screen capture module acquires full-screen frames and removes the occlusion of the program’s own window via Win32gui technology to ensure detection accuracy. The YOLOv8 object detection module identifies UI elements from screenshots and outputs their categories, confidence scores and bounding box coordinates. The mouse event monitoring module built on the Pynput library acts as the trigger for the entire detection pipeline, initiating capture and recognition only upon user clicks to reduce unnecessary resource consumption. The image annotation and storage module marks non-compliant elements with red bounding boxes and standard 44×44 pixel reference frames, then preserves the evidence locally. The data preprocessing module performs format conversion and normalization to standardize model input. The central scheduling module coordinates the execution sequence and data flow across all modules. The risk assessment module captures network requests through mitmproxy and evaluates the security level of redirected links based on multi-dimensional features such as URL structure and domain legitimacy.
    \item \textit{Interaction layer}: This layer acts as the interactive medium between users and the underlying program logic, delivering a graphical user interface with both information display and functional control components. It receives user instructions, presents real-time operating status and structured detection results, and overall improves the usability and observability of the system.
\end{enumerate}

\begin{figure}[H]
  \centering
  \begin{tikzpicture}[
  font=\small,
  flow/.style={-{Latex[length=2mm]}, draw=AcademicInk!70, line width=0.8pt},
  module/.style={draw=AcademicInk!70, line width=0.8pt, rounded corners=3pt,
    fill=AcademicBlue!7, align=center, minimum width=38mm, minimum height=22mm}
]
  \node[module] (data) at (-4.6,0) {\textbf{Data layer}\\Image storage\\Operation logs};
  \node[module] (logic) at (0,1.2) {\textbf{Business logic layer}\\Format conversion\\Image recognition\\Data processing};
  \node[module] (interaction) at (4.6,0) {\textbf{Interaction layer}\\User controls\\Event feedback\\Result statistics};

  \draw[flow] (data.north east) to[out=35,in=180] (logic.west);
  \draw[flow] (logic.east) to[out=0,in=145] (interaction.north west);
  \draw[flow] (interaction.south west) to[out=-145,in=-35] (data.south east);
\end{tikzpicture}
  \caption{Data, business-logic, and interaction layers of the proposed system.}
  \label{fig:modules}
\end{figure}
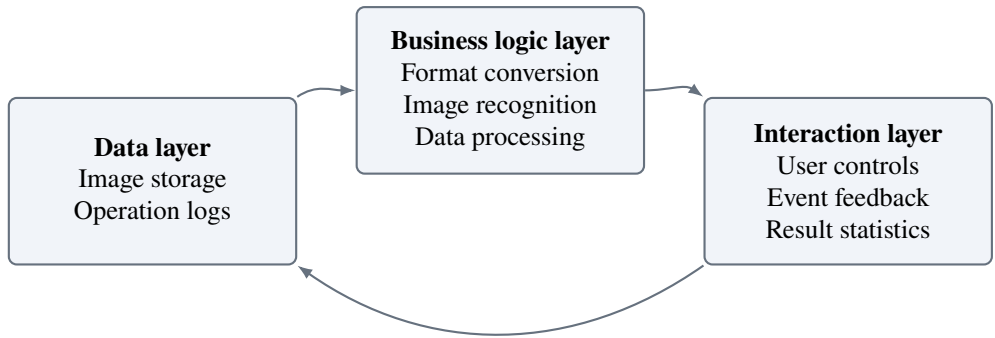

The graphical user interface of the system is composed of five distinct information regions, each with dedicated functions to support intuitive operation and result observation:

\begin{enumerate}[label=(\arabic*)]
  \item Real-time preview area, occupying approximately one-third of the main window, displays the processed clean screen capture with the program’s own window removed, allowing users to verify image clarity, check for target element occlusion, and adjust window position accordingly to guarantee detection integrity.
  \item Risk-prediction panel, arranged adjacent to the preview area, presents the security assessment results of captured network requests, including risk level, domain information and request method, enabling users to intuitively judge the safety of page redirections.
  \item Operation log area, located in the lower-right section, displays detection records in reverse chronological order with the latest entries prioritised, and supports up to 14 recent entries displayed simultaneously in half-screen mode, facilitating users to trace the detection process and review historical results.
  \item JSON panel, showing parsed packet data from captured network traffic, automatically truncates redundant content while retaining core fields for excessively large payloads, ensuring efficient information presentation without causing interface clutter.
  \item Summary statistics panel, aggregating all recorded operation and detection data, provides a holistic overview of total detections, violation counts and risk distribution for comprehensive analysis.
\end{enumerate}

Four control buttons are deployed at the top of the interface, respectively for activating visual recognition, switching recognition modes, toggling network monitoring, and pausing the detection process. Two configurable recognition modes are available: the precise detection mode focuses on size compliance judgment of the clicked element based on the 44×44 pixel threshold, while the full-scene recognition mode identifies and marks all detectable UI elements on the current screen, adapting to different inspection scenarios. Notably, the visual detection module and network monitoring module can be activated independently with separate evaluation criteria, allowing users to flexibly configure functions according to actual testing requirements.

Benefiting from the modular and layered architecture, the system achieves clear responsibility separation and low coupling between modules. For example, the detection accuracy can be improved by optimizing the YOLOv8 model in the business logic layer without modifying the underlying storage mechanism, while alternative storage backends can be integrated by only adjusting the data layer. This design not only simplifies the development and debugging process, but also reserves sufficient space for subsequent function expansion and performance optimization.

\subsection{Interaction Logic among Functional Modules}

Modules cooperate with each other via data transmission and event‑driven invocation, constructing a complete end‑to‑end detection workflow. When the system launches, the central scheduler initialises all functional components sequentially: it loads the pre‑trained YOLOv8 model, starts the screen capture thread for real‑time preview rendering, and activates the mouse event listener implemented by Pynput. After initialisation, the system enters an idle state and waits for user input events. Once a mouse click event is captured, the listener obtains the corresponding coordinate information and invokes PyAutoGUI to capture a full‑screen snapshot. If needed, Win32gui will be utilised to mask the window area of the detection tool itself on the screenshot. The processed image is then delivered to the YOLOv8 detection module. The detector returns multiple recognised UI objects, each attached with category label, confidence value and bounding box coordinates. The data processing unit matches the click coordinate with all detected UI elements to find the closest icon or button, and compares its width and height with the \(44\times44\) CSS‑pixel compliance threshold. If the clicked target is a close or cancel button whose size fails to reach the standard value, the violation handling process will be activated. The program draws a red bounding box and a red dot to mark the non‑compliant element, attaches a standard \(44\times44\) green reference frame below the target, and adds text annotations containing violation type and actual dimensions. The annotated screenshot will be stored in the assigned local directory with a timestamp as its filename, and this operation record will be synchronously written to the log panel together with updated statistical data. Two detection modes are available in this system: \textit{precision mode} as the default scheme only inspects the UI element at the clicked position, while \textit{full‑screen mode} identifies and categorises all visible interface components. Users can enable these two modes simultaneously to cross‑check detection outcomes. The program occupies nearly no CPU resources under idle conditions, and computation overhead only rises temporarily during model inference.\par
Users may manually turn on the mitmproxy‑based traffic capture module for network risk evaluation. After activation, all outbound HTTP and HTTPS requests will be forwarded to the proxy. The system filters valid traffic through a pre‑configured trusted domain whitelist and extracts the risk feature set introduced in Section 2.6. These extracted features are input into a machine learning classifier to generate a risk score, which is displayed on the dedicated risk prediction panel. The network monitoring module works as an independent optional component; users can execute UI visual inspection, traffic risk analysis, or both functions simultaneously as required. All detection outputs including UI compliance labels and network risk scores are synchronously displayed on the graphical interface in real time, which helps users rapidly locate risky interactive elements and adopt corresponding preventive measures.

\section{Experiments and Evaluation}
\label{sec:evaluation}

\subsection{Experimental Environment}

The hardware configuration adopted an Intel Core i7-12500H processor, 16~GB DDR4 memory, an NVIDIA GeForce GPU, a 512~GB SSD, and a $1920\times1080$-pixel display. The software environment was based on Windows and Python, with core libraries including YOLOv8, PyAutoGUI, Pynput, Win32gui, tkinter, Pillow, and mitmproxy. The YOLO training data consisted of three self-built datasets and one public dataset, containing 100, 150, 150, and 500 images respectively. The image corpus covered office applications, web pages, games, and payment-oriented Mini-Program screens. The self-built datasets contained 20, 12, and 22 labelled object categories, while the public dataset defined seven target classes. For network risk assessment, we constructed a dataset of 1,000 URL samples characterised by the seven features summarised in \cref{tab:url-features}.

\subsection{Testing Procedure and Error Analysis}

We evaluated multiple open-source libraries and tools for screen capture, visual detection, and mouse event monitoring to validate the performance and stability of the proposed system under diverse scenarios.
We first combined OpenCV with PyAutoGUI to implement screen capture and basic visual recognition. However, this combination failed to achieve satisfactory detection accuracy. Three major limitations are summarised as follows:
\begin{enumerate}[label=(\arabic*)]
    \item Unstable text extraction sensitive to capture boundaries. Text recognition quality heavily depended on the cropped region size. Small capture windows could not fully contain target elements, whereas oversized frames introduced redundant background noise and garbled text outputs. Disturbances such as cursor occlusion, bounding box scaling, and background graphics further degraded robustness; advertising graphics were frequently misclassified as text blocks.
    \item Low-level classification rules lacking semantic understanding. The pipeline only leveraged contour shape, component size, and pixel colour as discriminative features without high-level semantic reasoning. For complex interactive UI widgets, it merely generated coarse labels such as \textit{rectangular block} or \textit{yellow graphic}, and could not identify the functional semantics of each element.
    \item Static position heuristics causing frequent false positives. Detection relied on hard-coded spatial assumptions: for example, marking the top-right area as a close-button zone, the central region as a commodity panel, and the bottom area as a navigation bar. Even when no interactive elements existed in these predefined regions, the algorithm still produced invalid detection records.
\end{enumerate}

We further investigated Airtest as a cross-platform auxiliary tool for screen acquisition and visual testing. Airtest performs well in capturing mobile screens via USB connections and supports multi-platform deployment. Nevertheless, severe integration obstacles emerged when embedding it into our detection framework. The coordinate outputs exported by Airtest could not be stably parsed; the extracted position information actually originated from the underlying PyAutoGUI interface rather than Airtest’s native detection module. In addition, repeated activation of the startup routine spawned redundant Airtest processes, making it impossible to trace which instance generated valid detection results.

Considering the drawbacks of the above two schemes, we adopt YOLOv8 as the core object detection model in this work. Although YOLOv8 cannot directly conduct text recognition, it delivers competitive performance in graphic classification and precise prediction of element coordinates and dimensions, which perfectly satisfies the requirements of UI button and icon compliance inspection. We carried out comparative experiments over multiple datasets and hyperparameter combinations to determine the optimal training configuration. Four datasets were constructed for evaluation:
\begin{itemize}[label=\textbullet]
    \item Custom Dataset 1: Focused on shopping Mini-Program interfaces with 20 object categories;
    \item Custom Dataset 2: Containing 150 screenshots of diverse complex scenes with 13 generalised labels;
    \item Custom Dataset 3: Including 150 samples with evenly distributed element categories;
    \item Public Dataset: Composed of 500 generic interface images covering seven types of standard UI controls.
\end{itemize}

\subsection{Comparative Performance}

\subsubsection{Screen-Capture Functions}

We benchmarked PyAutoGUI’s \texttt{screenshot()} and Win32gui-based \texttt{BitBlt} under $1920\times1080$ resolution. \cref{tab:capture} records execution latency, CPU usage, memory consumption, perceived lag, and visual clarity. The \texttt{BitBlt} routine achieves faster screenshot acquisition and lower overall resource consumption. However, the practical differences in display clarity and interface stuttering are barely perceptible during real-world testing. A critical limitation of \texttt{BitBlt} is its exclusive support for Windows systems, while \texttt{screenshot()} enables cross-platform deployment over Windows, macOS, and Linux. Since our prototype prioritises multi-system compatibility and the performance gap between the two approaches remains modest, we adopt \texttt{screenshot()} as the core screen capture method.

\begin{table}[t]
  \centering
  \caption{Comparison of screen-capture functions.}
  \label{tab:capture}
  \begin{tabular}{@{}lrrrrl@{}}
    \toprule
    \textbf{Function} & \textbf{Time (ms)} & \textbf{CPU (\%)} & \textbf{Memory (MB)} & \textbf{Lag} & \textbf{Clarity} \\
    \midrule
    \texttt{screenshot()} (PyAutoGUI) & 193.5 & 5.1 & 9.1 & Slight & Good \\
    \texttt{BitBlt} (Win32gui)        & 124.7 & 4.3 & 8.2 & Slight & Good \\
    \bottomrule
  \end{tabular}
\end{table}

\subsubsection{Object-Detection Models}

YOLOv8 and YOLOv7 were evaluated under identical hardware conditions \cite{wang2023yolov7}and trained on the same dataset. Quantitative results are presented in \cref{tab:model}. YOLOv8 outperforms YOLOv7 across all metrics, with substantial improvements in precision and recall. Training hyperparameters such as batch size and epoch count exert significant influence on final performance: insufficient epochs lead to near-zero accuracy, whereas excessive training iterations may induce performance degradation. Training duration scales linearly with dataset size. Under CPU-only training, 100 images trained over 100 epochs required approximately 1.5 hours; by contrast, training on 500 images for the same epoch setting consumed nearly 4 hours.

\begin{table}[t]
  \centering
  \caption{Reported performance of the YOLO models.}
  \label{tab:model}
  \begin{tabular}{@{}lrrrr@{}}
    \toprule
    \textbf{Model} & \textbf{Inference time (ms)} & \textbf{Precision (\%)} & \textbf{Recall (\%)} & \textbf{F1 (\%)} \\
    \midrule
    YOLOv8 & 116.9 & 69.8 & 66.7 & 68.2 \\
    YOLOv7 & 123.4 & 50.6 & 49.5 & 55.6 \\
    \bottomrule
  \end{tabular}
\end{table}

\subsubsection{Dataset Comparison}

\begin{figure}[H]
  \centering
  \resizebox{0.5\textwidth}{!}{\includegraphics{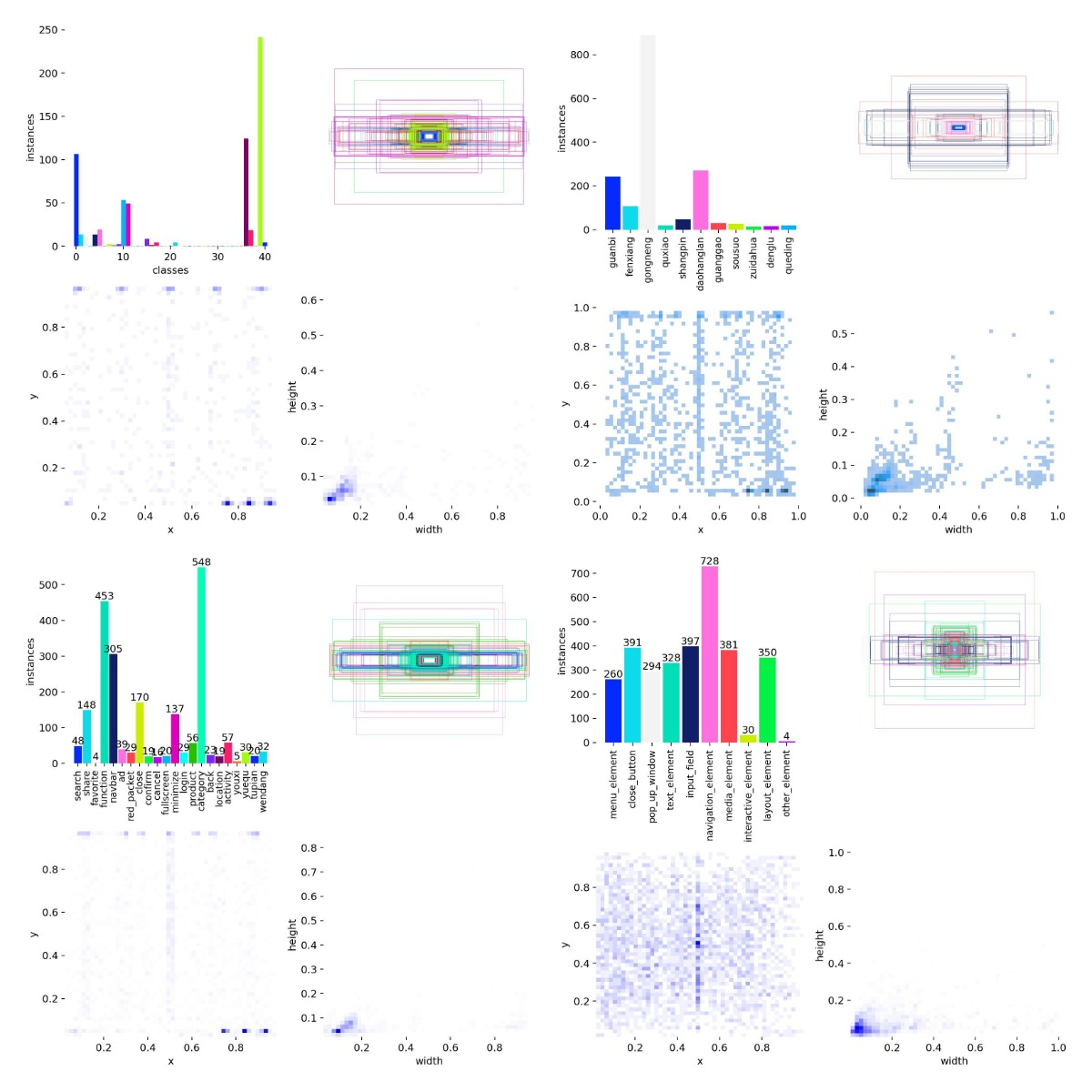}}
  \caption{Label distribution of each dataset.}
  \label{fig:dataset-dist}
\end{figure}

Models trained on different datasets exhibit distinct detection performance. It is inappropriate to judge the practical value of a dataset purely through isolated quantitative metrics. The label distribution and statistical indicators of the four datasets are visualised in \cref{fig:dataset-dist}, followed by analysis of real-world detection capability.

We evaluated detection performance on all four datasets. Quantitative results are listed in \cref{tab:datasets}.

\begin{table}[H]
  \centering
  \caption{Reported training results for the four datasets.}
  \label{tab:datasets}
  \begin{tabular}{@{}lrrrrr@{}}
    \toprule
    \textbf{Dataset} & \textbf{Precision (\%)} & \textbf{Recall (\%)} & \textbf{F1 (\%)} & \textbf{CPU (\%)} & \textbf{Memory (\%)} \\
    \midrule
    Custom Set 1 & 69.7 & 47.4 & 56.5 & 5.1 & 9.1 \\
    Custom Set 2 & 69.8 & 66.7 & 68.2 & 2.1 & 5.5 \\
    Custom Set 3 & 98.8 & 99.0 & 99.2 & 1.4 & 7.0 \\
    Public Set 4 & 63.0 & 54.0 & 58.2 & 1.8 & 5.4 \\
    \bottomrule
  \end{tabular}
\end{table}

Custom Set~1 contained limited samples and mainly covered shopping and food-delivery interfaces, resulting in weak generalisation ability. In Custom Set~2, product widgets occupied more than half of all annotations, introducing bias toward over-prediction of this category. Custom Set~3 featured balanced category distribution and achieved the best practical detection performance, with the only exception of rare controls such as maximise and share buttons. Public Set~4 contained more samples but was not tailored to Mini-Program UI scenarios, leading to the lowest practical detection accuracy. Therefore, Custom Set~3 was selected for all subsequent experiments.

\subsection{Comparison of Packet-Capture Tools}

We compare mitmproxy with two mainstream traffic analysis tools, Fiddler and Wireshark, in terms of integration difficulty, HTTPS decryption capability, filtering flexibility, and output readability. Fiddler can capture and parse Mini-Program domain names and JSON payloads, extracting identifiers, style parameters, titles, and other critical fields. However, HTTPS decryption requires manual installation of third-party certificates. Its filtering mechanism is rigid: target domains must be configured line-by-line, and rules cannot be globally applied to all WeChat processes. On the positive side, parsed data is well-structured and easy for developers to inspect. 

Wireshark captures all network traffic and classifies packets by protocol (DNS, HTTP, ARP, UDP, etc.), yet it cannot decrypt HTTPS payloads by default. Decryption requires private key configuration, which is less convenient than certificate-based solutions. Filtering valid Mini-Program traffic is time-consuming, and outputs mainly consist of domain names and ciphertext without structured application-layer information.

Mitmproxy provides comparable analytical power to Fiddler while supporting seamless integration into our Python-based framework. It supports flexible filtering rules to constrain captured traffic scope. Its primary drawback is the requirement to configure a system proxy, which may temporarily interrupt network access for other applications during interception. Despite this limitation, we adopt mitmproxy due to its native programmability and sufficient analytical capacity for our risk assessment module.

\subsection{Experimental Findings and Optimisation Directions}

Three core conclusions are derived from comprehensive testing. First, GPU hardware performance and library versions directly affect object detection latency. Frequent full-screen capture at short sampling intervals via PyAutoGUI causes severe system stalling. Second, dataset quality and domain whitelist rules dominate overall detection accuracy: high-quality annotated datasets improve model performance, while refined domain filtering effectively eliminates irrelevant network traffic noise. Third, parameter mismatches between the capture and detection modules lead to stuttering, crashes, or functional failure; consistent parameter tuning across all modules is essential for system stability.

Detection bounding boxes occasionally deviate from actual UI boundaries when background colours are highly similar to target widgets. Tiny hidden close buttons are frequently occluded by large commodity panels, triggering misjudgements. For network parsing, the built-in JSON viewer outputs complete raw data but lacks hierarchical classification like Fiddler, and lengthy request payloads cannot be fully displayed. Accordingly, URL domain compliance serves as the core risk indicator, while parsed JSON content is only used as auxiliary evidence.

Four improvement directions are summarised to address the identified limitations:
\begin{enumerate}[label=(\arabic*)]
    \item \textit{Detection Performance Upgrade}: Expand annotated datasets and tune YOLOv8 hyperparameters (learning rate, batch size) with data augmentation to boost small-object detection; adjust the size threshold to reduce false negatives. Static code analysis may be integrated to complement visual inspection, though additional computation may sacrifice real-time responsiveness.
    \item \textit{Runtime Efficiency Boost}: Adopt faster screen-capture APIs and multi-thread parallelism for screenshot acquisition, preprocessing, and evidence storage. Redundant code pruning and dynamic resource allocation will lower CPU and memory overhead for smoother execution.
    \item \textit{Functional Expansion}: Implement batch detection, CSV/Excel result export, adjustable image compression, and multi-window inspection modules to support large-scale testing scenarios.
    \item \textit{User Experience Enhancement}: Simplify GUI workflows with detailed operation prompts, customisable detection ranges and target categories. The error feedback module will provide actionable advice for non-compliant UI elements together with problem-report channels. An auto-update pipeline will support manual labelling of misdetected icons for iterative model fine-tuning.
\end{enumerate}

Full implementation of the above optimisations will enhance the accuracy, speed, usability, and extensibility of the tool for complex Mini-Program testing scenarios.

\section{Conclusion}
\label{sec:conclusion}

This work develops an integrated detection prototype for WeChat Mini-Programs, combining YOLOv8-based visual UI inspection and mitmproxy traffic analysis to assess UI compliance and predict network risks. Traditional single-modal approaches relying on colour matching or template matching degrade significantly under variable icon scales and lighting conditions. The proposed hybrid vision-traffic framework overcomes this limitation by cross-verifying element geometric attributes and URL security features. Configurable filtering rules balance data completeness and processing efficiency to realise flexible risk evaluation. The designed pipeline accurately locates, measures, and archives non-compliant interactive widgets. With extended training data and parameter tuning, the system generalises across diverse Mini-Program interfaces and reveals hidden threats behind seemingly standard UI components. It delivers a standardised testing scheme to regulate Mini-Program UI design, protects users’ property interests, and promotes healthy development of the Mini-Program ecosystem.

Four promising research directions are outlined based on current prototype limitations and industry trends\cite{al2025adaptive,che2025uncovering,xiang2026minieval,baskaran2023measuring,bari2025filter,2024Sybil,2025DoS,2025A}. The first direction concerns diversified security detection. It requires refining YOLOv8 to target Mini-Program-specific threats such as malicious pop-ups, unauthorised access entrances, and deceptive redirect links, as well as integrating NLP techniques to identify risky embedded text content. The second direction focuses on developing a full-lifecycle monitoring system. An end-to-end monitoring platform can be constructed to support real-time network metric tracking and instant risk alerts. Log mining will also be adopted to continuously dig out latent security vulnerabilities hidden in Mini-Program operation processes. The third direction is cross-platform intelligent iteration. The existing Windows-only framework can be migrated to adapt to Linux and macOS systems. Besides, reinforcement learning algorithms such as Q-Learning can be introduced to achieve adaptive model optimisation. Cloud storage will be deployed to realise the cross-device synchronisation of screenshots and logs, while voice and gesture control functions can be incorporated to lower the cost of manual operations. The last direction targets interconnected security defense. Data interoperability with mainstream Mini-Program security tools can be enabled to support the automatic interception of malicious links and illegal content. Further efforts can be made to explore lightweight algorithms and native Mini-Program test clients to implement portable inspection on mobile terminals. Continuous cross-domain research in these fields is expected to promote the development of Mini-Program security research and deliver reliable security protection for massive end users.
\\

\section*{Acknowledgments}
AI-based tools are used for language polishing during manuscript preparation.

\bibliography{references}

\end{document}